\input{aipcheck.tex}
\documentclass[final]{aipproc}
\layoutstyle{6x9}
\def\be{\begin{equation}}
\def\ee{\end{equation}}
\def\bea{\begin{eqnarray}}
\def\eea{\end{eqnarray}}
\def\ba{\begin{array}}
\def\ea{\end{array}}

\begin{document}
\title[Electromagnetic form factors of the nucleon]{Electromagnetic form factors of the nucleon in the chiral constituent quark model}
\author{Harleen Dahiya}{address={Department of Physics, Dr. B.R. Ambedkar National Institute of Technology,
Jalandhar, Punjab-144 011, India.}}
\author{Neetika Sharma}{address={Department of Physics, Dr. B.R. Ambedkar National Institute of Technology,
Jalandhar, Punjab-144 011, India.}}

\date{\today}

\begin{abstract}

The electromagnetic form factors have attracted lot of theoretical and experimental attention recently as they encode extensive information on the internal structure of the hadron. An understanding of the form factors is necessary to describe the strong interactions as they are sensitive to the pion cloud and provide a test for the QCD inspired effective field theories based on the chiral symmetry. In view of the very exciting recent developments in the field, we propose to apply the techniques of chiral constituent quark model to measure the electromagnetic form factors of the nucleon.The results obtained are comparable to the latest experimental studies and also show
improvement over some theoretical interpretations.

\end{abstract}

\maketitle

\section{Introduction}

The knowledge of internal structure of
nucleon in terms of quark and gluon degrees of freedom in QCD
provides a basis for understanding more complex, strongly
interacting matter.  The electromagnetic form factors are the fundamental quantities of
theoretical and experimental interest to investigate the internal
structure of nucleon. Recently, a wide variety of accurately
measured data have been accumulated for the static properties of
baryons, for example, masses, electromagnetic moments, charge
radii etc.. which are important as they lie in the nonperturbative regime
of QCD. While QCD is
accepted as the fundamental theory of strong interactions, it
cannot be solved accurately in the nonperturbative regime. A
coherent understanding of the hadron structure in this energy
regime is necessary to describe the strong interactions as they
are sensitive to the pion cloud and provide a test for the QCD
inspired effective field theories based on the chiral symmetry. A
promising approach is offered by constituent-quark models which can be modified
to include the relevant properties of QCD in the nonperturbative regime, notably the
consequences of the spontaneous breaking of chiral symmetry
($\chi$SB).

\section{Electromagnetic form factors}

The internal structure of nucleon is determined in terms of
electromagnetic Dirac and Pauli from factors $F_1(Q^2)$ and
$F_2(Q^2)$ or equivalently in terms of the electric and magnetic
Sachs form factors $G_E(Q^2)$ and $G_M(Q^2)$ \cite{sach}. The
issue of determination of the form factors has been revisited in
the recent past with several new experiments measuring the form
factors with precision at MAMI \cite{mami} and JLAB \cite{jlab} which are in
significant disagreement with those obtained from the Rosenbluth
separation \cite{rosen1}. This inconsistency leads to a large
uncertainty in our knowledge of the proton electromagnetic form
factors and urge the necessity for the new parameterizations and
analysis \cite{fit}.

The most general form of the hadronic current for a spin
$\frac{1}{2}$-nucleon with internal structure is given as \be
\langle B| J^\mu_{{\rm had}}(0)|B \rangle = \bar{u}(p')
\left(\gamma^\mu F_1(Q^2)+ i{\sigma^{\mu\nu}\over 2M}q_\nu
F_2(Q^2)\right) u(p), \label{ff} \ee where $u(p)$ and $u(p')$ are
the 4-spinors of the nucleon in the initial and final states
respectively. The Sachs form
factors $G_E$ and $G_M$ can be related to the Dirac and Pauli form
factors and the  Fourier transform can be expressed
in terms of the nucleon charge density. The most general form of the
multipole expansion in the
spin-flavor space is
\be{\mathrm A'} \sum_{i=1}^3 e_i {\bf 1} - {\mathrm B'}
\sum_{i \ne j}^3 e_i \bigg[ 2 {\bf{\sigma}_i} \cdot
\bf{\sigma}_j - (3{\bf{\sigma}_{iz}} {\bf{\sigma}_{jz}} -
\bf{\sigma}_i \cdot \bf{\sigma}_j) \bigg] - {\mathrm C'}
\sum_{i \ne j \ne k}^3 e_i \bigg[ 2 {\bf{\sigma}_j} \cdot
\bf{\sigma}_k- (3 {\bf{\sigma}_{jz}} {\bf{\sigma}_{kz}} -
\bf{\sigma}_j \cdot \bf{\sigma}_k) \bigg] \,.\ee
The charge
radii operator composed of one-, two-, and
three-quark terms is expressed as
\[ \widehat{r^2} = {\mathrm A} \sum_{i=1}^3 e_i {\bf 1} +
{\mathrm B} \sum_{i \ne j}^3 e_i \, {\bf{\sigma}_i} \cdot
\bf{\sigma}_j + {\mathrm C} \sum_{i \ne j \ne k }^3 e_i \,
\bf{\sigma}_j \cdot \bf{\sigma}_k \,, \] whereas the
quadrupole moment operator can be expressed as \[ \widehat{Q} = {\mathrm B'} \sum_{i
\ne j}^3 e_i \left( 3 \sigma_{i \, z} \sigma_{ j \, z} -
\bf{\sigma}_i \cdot \bf{\sigma}_j \right) + {\mathrm C'} \!\!
\sum_{i \ne j \ne k }^3 e_i \left( 3 \sigma_{j \, z} \sigma_{ k
\, z} - \bf{\sigma}_j \cdot \bf{\sigma}_k \right) \,.  \] The
coefficients ${\mathrm A} = {\mathrm A'}$, ${\mathrm B}= -2
{\mathrm B'}$, and ${\mathrm C} = -2 {\mathrm C'}$ are the general
parameterization (GP) method parameters \cite{morp89,buch2002}.
\section{charge radii}

The mean square charge radius ($r^2_B$), giving the possible
``size'' of baryon, has been investigated experimentally with the
advent of new facilities at JLAB, SELEX Collaborations
\cite{jlab2,pdg}. Several measurements have been made for the
charge radii of $p$, $n$, and ${\Sigma^-}$ in electron-baryon
scattering experiments \cite{sigma-,wa89} giving $r_p = 0.877 \pm
0.007$ fm ($r^2_p = 0.779 \pm 0.025$ fm$^2$ \cite{rosenfelder}) and
$r^2_n = -0.1161 \pm 0.0022$ fm$^2$ \cite{pdg}.

The charge radii operators for the spin $\frac{1}{2}^+$ octet and
spin $\frac{3}{2}^+$ decuplet baryons can be expressed in terms of the flavor $(\sum_i e_{i})$ and spin $(\sum_i
e_i \sigma_{iz})$ structure of a given baryon as \bea
\widehat{r^2_{B}} &=&  ({{\mathrm A}}- 3{\mathrm B})\sum_i e_i +
3( {\mathrm B} - {\mathrm C}) \sum_i e_i \sigma_{i z} \,,
\label{r1/2}
\\ \widehat{ r^2_{B*}} &=& ({\mathrm A} - 3{\mathrm B} + 6 {\mathrm
C}) \sum_i e_i + 5( {\mathrm B} - {\mathrm C}) \sum_{i} e_i
\bf{\sigma_{iz}} \,.\label{r3/2} \eea It is clear from the above
equations that the determination of charge radii basically reduces
to the evaluation of the . The charge radii
squared ${r^2_{B(B^*)}}$ for the octet (decuplet) baryons can now be
calculated by evaluating matrix elements corresponding to the
operators in Eqs. (\ref{r1/2}) and (\ref{r3/2}) and are given as $ r^2_B = \langle B |\widehat{ r^2_B}| B \rangle$, $r^2_{B^*} = \langle B^* |\widehat{r^2_{B^*}}| B^* \rangle$.
Here, $|B \rangle$ and $|B^* \rangle$
respectively, denote the spin-flavor wavefunctions for the spin
$\frac{1}{2}^+$ octet and the spin $\frac{3}{2}^+$ decuplet baryons.

The naive quark
model (NQM) \cite{nqm} calculations show that the results are in disagreement
with the available experimental data. In this context, the chiral constituent quark
model ($\chi$CQM) \cite{manohar,hd}, which incorporates chiral symmetry breaking, has been extended to calculate the charge radii of spin $\frac{1}{2}^+$ octet and
spin $\frac{3}{2}^+$ decuplet baryons using GP method. A
redistribution of flavor and spin takes place among the ``sea
quarks'' in the interior of hadron due to the fluctuation process
and chiral symmetry breaking  in the $\chi$CQM.
The most significant prediction of
the model is the non-zero value pertaining to the charge radii of
the neutral octet baryons ($n$, $\Sigma^0$, $\Xi^0$, and
$\Lambda$) and decuplet baryons ($\Delta^0$, $\Sigma^{*0}$,
$\Xi^{*0}$). The effects of  SU(3) symmetry breaking have also
been investigated and the results show considerable improvement
over the SU(3) symmetric case. New experiments aimed at measuring the charge radii of the other
baryons are needed for a profound understanding of the hadron
structure in the nonperturbative regime of QCD.

\section{quadrupole moments}

Recent experimental developments \cite{kelly,glashow}, providing
information on the radial variation of the charge and magnetization
densities of the proton, give the evidence for a deviation of the
charge distribution from spherical symmetry. Since
the quadrupole moment of the nucleon should vanish on account of its spin-1/2 nature, this observation has naturally turned to be the
subject of intense theoretical and experimental activity. In this
context, $\Delta$(1232) resonance being the lowest-lying excited
state of the nucleon, plays a very important role in the low energy
baryon phenomenology.

The spin and parity selection rules in the $\gamma + p \to \Delta^+$
transition allow three contributing amplitudes, the magnetic dipole
$G_{M1}$, the electric quadrupole moment $G_{E2}$, and the charge
quadrupole moment $G_{C2}$ photon absorption amplitudes \cite{tia03,blanpied}. The
$G_{M1}$ amplitude gives us information on magnetic moment whereas
the information on the intrinsic quadrupole moment can be obtained
from the measurements of $G_{E2}$ and $G_{C2}$ amplitudes
\cite{pdg}. If the charge distribution of the initial and final
three-quark states were spherically symmetric, the $G_{E2}$ and
$G_{C2}$ amplitudes of the multipole expansion would be zero
\cite{becchi}. However, recent results on non-zero quadrupole amplitudes \cite{jlab2,ber03} lead to the conclusion that the nucleon and the
$\Delta^+$ are {\it intrinsically} deformed.

For the case of octet baryons we find that
the quadrupole moments are zero for all the cases in NQM. In the SU(3) symmetry breaking limit,
the ``small'' numeric value of quadrupole moment measures the deviation in shape of baryons from the spherical symmetry. The predicted signs of {\it intrinsic} quadrupole
moment are important as they measure the type of deformation in
the baryon. The small observed negative value of $p$ and $n$ quadrupole moments suggest that these are oblate in shape which is in agreement
with several other calculations in literature.

For the case
of decuplet baryons, the quadrupole moments of the charged baryons
are equal whereas all neutral baryons have zero quadrupole moment.
The results in NQM using the GP method
predict an oblate shape for all positively charged baryons
($\Delta^{++}$, $\Delta^+$, and $\Sigma^{*+}$), prolate shape for
negatively charged baryons ($\Delta^-$, $\Sigma^{*-}$, $\Xi^{*-}$,
and $\Omega^-$). On incorporating the effects of
chiral symmetry breaking and ``quark sea'' in the $\chi$CQM, a small
amount of prolate deformation in neutral baryons ($\Delta^0$,
$\Sigma^{*0}$, and $\Xi^{*0}$) is observed.

\section{ACKNOWLEDGMENTS}
HD would like to thank the organizers of DIS2011 for financial support. This work was partly supported by Department of
Atomic Energy, Government of India through Grant No.
2010/37P/48/BRNS/1445.

\end{document}